\documentclass[11pt]{article}
\usepackage{latexsym}  
\usepackage{amssymb}
\usepackage{graphicx}
\usepackage{amsmath}

\begin{document}
\hspace*{12cm} {MISC-2012-10}

\begin{center}
{\Large\bf Family Gauge Bosons with an Inverted Mass Hierarchy\footnote{
Talk presented by Y.~Koide at International Workshop on Grand Unified 
Theories (GUT2012), March 15-17, 2012 at YITP, Kyoto, Japan. 
To appear in the proceedings which will be published by AIP.  
}
}

\vspace{4mm}
{\bf Yoshio Koide$^{a,b}$ and Toshifumi Yamashita$^{b}$}

${}^{a}$ {\it Department of Physics, Osaka University, 
Toyonaka, Osaka 560-0043, Japan} \\
{\it E-mail address: koide@kuno-g.phys.sci.osaka-u.ac.jp}

${}^b$ {\it   MISC, 
Kyoto Sangyo University,  
Kyoto 603-8555, Japan}\\
{\it E-mail address: tyamashi@cc.kyoto-su.ac.jp}

\end{center}

\vspace{3mm}
\begin{abstract}
A model that gives family gauge bosons with an inverted 
mass hierarchy is proposed, stimulated by Sumino's cancellation 
mechanism for the QED radiative correction to the charged 
lepton masses.
The Sumino mechanism cannot straightforwardly be applied to SUSY 
models because of the non-renormalization theorem.   
In this talk, an alternative model which is applicable 
to a SUSY model is proposed.
It is essential that family gauge boson masses $m(A_i^j)$ 
in this model is given by an inverted mass hierarchy 
$m(A_i^i) \propto 1/\sqrt{m_{ei}}$, in contrast to 
$m(A_i^i) \propto \sqrt{m_{ei}}$ in the original Sumino model.
Phenomenological meaning of the model is also investigated.  
In particular, we notice a deviation from the $e$-$\mu$ universality in the
tau decays.
\end{abstract}

\vspace{3mm}

\section{ Motive to consider an inverted mass hierarchy}

In this section, it is discussed why we consider a family gauge 
boson model with an inverted mass hierarchy. 
However, since our model has been stimulated by a Sumino mechanism,
prior to giving our motive, we would like to give a brief review
of the Sumino mechanism.  

\subsection{ Why did Sumino need a family gauge symmetry?}

In 2009, Sumino \cite{Sumino09PLB} has seriously taken 
why the mass formula \cite{Koidemass}
\begin{equation}
K \equiv 
\frac{m_e +m_\mu +m_\tau}{(\sqrt{m_e} +\sqrt{m_\mu}
+\sqrt{m_\tau})^2} = \frac{2}{3} ,
\label{K-relation}
\end{equation}
is so remarkably satisfied with  the pole masses:
$ K^{pole}=(2/3)\times (0.999989\pm 0.000014)$.  
However, in a mass matrix model, usually, 
 ``masses" do not mean ``pole masses",  but ``running masses". 
The formula  \eqref{K-relation} is only valid with 
the order of $10^{-3}$ for the running masses, 
e.g. $K(\mu)=(2/3)\times (1.00189 \pm 0.00002)$ 
at $\mu =m_Z$. 
The deviation of $K(\mu)$ from $K^{pole}$ is 
caused by a logarithmic term $m_{ei}\log(\mu/m_{ei})$ 
in the QED radiative correction term \cite{Arason} 
\begin{equation}
m_{ei}(\mu) = m_{ei}^{pole} \left[ 1-\frac{\alpha_{em}(\mu)}{\pi}
\left(1 +\frac{3}{4} \log \frac{\mu^2}{m_{ei}^2(\mu)} \right)
\right].
\label{QED_mass}
\end{equation}

Meanwhile, under the transformation 
$m_i \rightarrow m_i (1+ \varepsilon_0 +\varepsilon_i)$, 
where $\varepsilon_0$ and $\varepsilon_i$ are family-number
independent and dependent factors, respectively, 
the ratio $K$ is invariant if $\varepsilon_i=0$. 
That is, $\varepsilon_i$ term corresponds to the $\log m_{ei}$
term in \eqref{QED_mass}.  
If we can remove the  $\log m_{ei}$ term, we can 
build a model with $K^{pole}=K(\Lambda)$. 
Therefore, Sumino \cite{Sumino09PLB} has proposed an idea that 
the $\log m_{ei}$ term is canceled by a contribution from family 
gauge bosons as shown in Fig.1. 

\begin{figure}[h]
\begin{picture}(400,50)(0,-20)
\put(10,-5){$\psi_L$}
\put(60,-5){$\psi_R$}
\put(30,-15){$m_i(\mu)$}
\put(150,-15){$m_i^{pole}$}
\put(255,-15){$e^2 \log m_{ei}^2$}
\put(380,-15){$-g_F^2 \log M_{ii}^2$}
\includegraphics[height=.05\textheight]{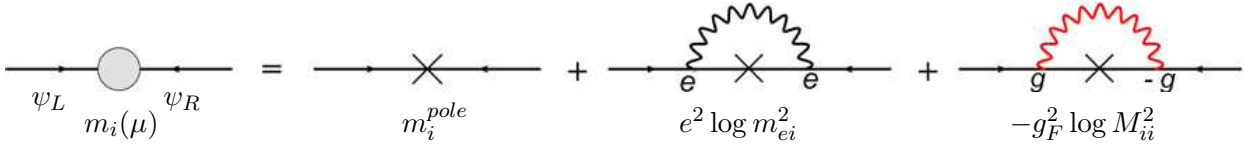}
\end{picture}  

  \caption{The Sumino mechansm}
\end{figure}

\vspace{3mm}

In order to work the Sumino mechanism correctlyC
the following conditions are essential:
(i)  the left- and right-handed
charged leptons $e_{Li}$ and $e_{Ri}$ are assigned to ${\bf 3}$
and ${\bf 3}^*$ of a U(3) family symmetry, respectively;   
(ii) masses of the gauge bosons $A_i^j$ are given by
\begin{equation}
M_{ij} \equiv m(A_i^j) \propto \sqrt{ m_{ei} +m_{ej}}
\label{Sumino_g_mass}
\end{equation}
Thus, the contribution $-g_F^2 \log M^2_{ii}$ term from the family 
gauge bosons can cancel the $e^2 \log m^2_{ei}$ term. 

Now, we want to apply this mechanism to a SUSY model.  
However, note that the contribution given in Fig.~1 (b) is zero 
in a SUSY model because it is a vertex correction type.  
Therefore, the Sumino mechanism cannot apply to a SUSY model.

\begin{figure}
  \includegraphics[height=.2\textheight]{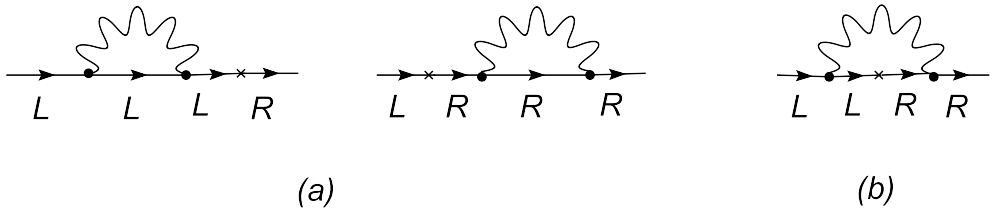}
  \caption{Contributions from family gauge bosons}
\end{figure}

\subsection{Our motive}

Here, we propose a Sumino-like mechanism which can work in a SUSY model. 
We notice the diagram (a) in Fig.~2.
\begin{equation}
\varepsilon_i = \rho \left( \log \frac{m_{ei}^2}{\mu^2} + \zeta \sum_j
\log \frac{M_{ij}^2}{\mu^2} \right) ,
\quad
\rho = \frac{3}{4} \frac{\alpha_{em}}{\pi} , \ \ \  
\zeta=\frac{2}{3} \frac{\alpha_F}{\alpha_{em}} ,
\label{epsilon_ei}
\end{equation}
In order that the cancellation works correctly, 
since $\zeta > 0$, we must consider
\begin{equation}
M_{ij}^2  \propto \frac{1}{m_{ei}} +   \frac{1}{m_{ej}} ,
\label{our_g_mass}
\end{equation}
differently from the Sumino's gauge boson mass 
relation \eqref{Sumino_g_mass}.

Note that since the gauge bosons $A_i^j$ with $j\neq i$ 
can contribute to the $\varepsilon_i$ term, differently 
from the Sumino model, the QED $\log m_{ei}$ term 
cannot be canceled 
by the gauge boson terms exactly. 
However, practically, the cancellation $R-1$ 
($R \equiv {K(m_{ei})}/{K(m^0_{ei})}$)
can be smaller than $10^{-5}$ at $\zeta=7/4$.

\section{Outline of the model}

By introducing two types of scalars $\Phi^\alpha_i$ 
(and $\bar{\Phi}^i_\alpha$)
and $\Psi^\alpha_i$ (and $\bar{\Psi}^i_\alpha$) 
which are $(3, 3^*)$ ($(3^*,3)$)
of family symmetries SU(3)$\times$SU(3)$'$,  
we build our model as follows:

\noindent
(i) The charged lepton mass matrix $M_e$ is given by
\begin{equation}
M_e \propto \langle \bar{\Phi} \rangle \langle \Phi \rangle .
\label{cond-1}
\end{equation}

\noindent
(ii) Family gauge boson masses $M_{ij}$ are dominantly
given by the VEV $\langle \Psi \rangle$. 
Therefore, we must show 
\begin{equation}
\langle \Psi \rangle \gg \langle \Phi \rangle .  
\label{cond-2}
\end{equation}

\noindent
(iii) Gauge bosons take an inverted mass hierarchy:
We must show 
\begin{equation}
\langle \Psi \rangle \langle \Phi \rangle \propto {\bf 1} .
\label{cond-3}
\end{equation}
The conditions (i), (ii) and (iii) are derived from the 
following superpotentials \cite{Koide-Yamashita_12}
\begin{equation}
W_Y = y_\ell \ell_i \bar{\Phi}^i_\alpha \bar{L}^\alpha
+ y_{Hd} {L}_\alpha H_d {E}^{c\alpha} + y_e \bar{E}^c_\alpha 
\Phi^\alpha_j e^{cj}
+ M_E E^{c\alpha}\bar{E}^c_\alpha 
+M_L L_\alpha  \bar{L}^\alpha ,
\label{W-PhiPhi}
\end{equation}
\begin{equation}
\begin{array}{l}
W_\Phi  = \lambda_1 \Phi^\alpha_i \bar{\Phi}^i_\alpha \theta_\Phi
-\lambda_2 S^2 \theta_\Phi , \\
W_{br} = \mu_S S \theta_S - \varepsilon \mu_S^2 \theta_S ,
\label{W-epsilon}
\end{array}
\end{equation}
\begin{equation}
\begin{array}{ll}
W_{\Phi\Psi}  = & \left( \lambda_A \bar{\Psi}^i_\alpha {\Phi}^\alpha_j
+ \bar{\lambda}_A \bar{\Phi}^i_\alpha {\Psi}^\alpha_j \right) 
(\Theta_A)^j_i 
+ \left(\lambda'_A \bar{\Psi}^i_\alpha \Phi^\alpha_i 
+ \bar{\lambda}'_A \bar{\Phi}^i_\alpha {\Psi}^\alpha_i
-\mu_A S \right) (\Theta_A)^j_j \\
 & + \left( \lambda_B {\Phi}^\alpha_i \bar{\Psi}^i_\beta 
+ \bar{\lambda}_B {\Psi}^\alpha_i \bar{\Phi}^i_\beta \right) 
(\Theta_B)^\beta_\alpha  
+ \left( \lambda'_B {\Phi}^\alpha_i \bar{\Psi}^i_\alpha
+ \bar{\lambda}'_B {\Psi}^\alpha_i \bar{\Phi}^i_\alpha
-\mu_B S \right) (\Theta_B)_\beta^\beta .
\end{array}
\label{W_PhiPsi}
\end{equation}
respectively.
As a result, we obtain
\begin{equation}
\langle {\Phi} \rangle = \langle \bar{\Phi} \rangle ={v}_{\Phi} {Z}, 
\ \ \ 
\langle {\Psi} \rangle = \langle \bar{\Psi} \rangle ={v}_{\Psi} {Z}^{-1},
\label{PhiPsiZ}
\end{equation}
where
\begin{equation}
Z={\rm diag}(z_1, z_2, z_3),
\label{Z}
\end{equation}
and the parameter values of $z_i$ are given by 
\begin{equation}
z_i =\frac{\sqrt{m_{ei}}}{\sqrt{ m_{e1}+m_{e2}+m_{e3}}} ,
\label{zi}
\end{equation}
where $(m_{e1}, m_{e2}, m_{e3})=(m_e, m_\mu, m_\tau)$. 
The explicit values of $z_i$ are given by
$(z_1,z_2,z_3) = (0.016473, 0.23688, 0.97140)$.
Thus, we can approximately estimate the family gauge boson masses 
$m(A^i_j)$ as follows
\begin{equation}
M_{ij}^2 \equiv m^2(A^i_j) \simeq 
 g_F^2 v_\Psi^2 \left( 
\frac{1}{z_i^2} + \frac{1}{z_j^2} \right)
\propto 
 \left(
\frac{1}{m_{ei}} + \frac{1}{m_{ej}} \right),
\label{Mij}
\end{equation}
if the mixing between the U(3) and U(3)$'$ gauge bosons can be neglected. 
This happens when the latter gauge bosons are sufficiently heavy. 

However, for reasons of my talk time, I would like to 
skip the details of the model. 
For more details, see our recent paper \cite{Koide-Yamashita_12}.

\section{How different from the Sumino model?}

Let us give only a summary table for differences between Sumino model
and ours in Table \ref{dif_S-KY}.

\begin{table}[h]
\begin{tabular}{lcc}
\hline
    & {Sumino model}
  & {Present model}  \\
\hline
 & non-SUSY & SUSY \\
U(3) assignment of $(e_L, e_R)$ & $\sim (3,3^*)$ & 
$\sim (3,3)$ \\
Anomaly &  a model with anomaly & an anomaly-less model \\
Gauge boson masses & Normal & Inverted \\
   & $M_{ij} \propto \sqrt{m_i + m_j}$
   & $M_{ij} \propto \sqrt{ \frac{1}{m_i}+ \frac{1}{m_j} }$ \\
Family currents & 
$(J_\mu)_i^j = \bar{\psi}_L^j \gamma_\mu \psi_{Li} 
- \bar{\psi}_{Ri} \gamma_\mu \psi_{R}^j $ & 
$(J_\mu)_i^j = \bar{\psi}^j \gamma_\mu \psi_{i}$ \\ 
Effective $\Delta N_f =2$ int. & appear even if $U_q ={\bf 1}$
& not appear in the limit of $U_q ={\bf 1}$ \\
\hline
\end{tabular}
\caption{Comarison between the Sumino model and the
present model.}
\label{dif_S-KY}
\end{table}

\section{How to observe the gauge boson effects}

Note that the family number is defined on the 
diagonal basis of the charged lepton mass matrix $M_e$. 
Hadronic modes are in general dependent on the quark
mixing matrices $U_u$ and $U_d$. 
We know the observed values of $V_{CKM} = U_u^\dagger U_d$,
but we do not know values of $U_u$ and $U_d$ individually. 
In the present model, we do not give an explicit model
for quark mixings on the diagonal basis of $M_e$. 
A constraint from $K^0$-$\bar{K}^0$ mixing is highly dependent
 on a model of $U_d$.  We will not discuss it in this talk. 

\subsection{Deviation from the $e$-$\mu$ universality in the tau decays}

The pure-leptonic decays are independent of a model of the quark mixing.
Therefore, first, we discuss the deviation from the $e$-$\mu$ universality 
in the tau decays.

We have family current-current interactions
\begin{equation}
\frac{G_{ij}}{\sqrt{2}} (\bar{\nu}_i \gamma_\mu \nu_j)
(\bar{e}_j \gamma^\mu e_i) ,
\label{family_exchange}
\end{equation}
where
${G_{ij}}/{\sqrt{2}} = {g_F^2}/{2 M_{ij}^2} \simeq {z_j^2}/{2 v_\Psi^2}$,
in addition to the conventional weak interactions
\begin{equation}
\frac{G_F}{\sqrt{2}} (\bar{e}_j \gamma_\mu (1-\gamma_5) \nu_j)
(\bar{\nu}_i \gamma^\mu (1-\gamma_5) e_i),
\label{W_exchange}
\end{equation}
where
${G_F}/{\sqrt{2}} = {g_W^2}/{8 M_W^2} = {1}/{2 v_W^2}$.

\begin{figure}[h]
\begin{picture}(500,70)(-120,0)
\put(-15,25){$\tau^-$}
\put(10,5){$A_3^2$ ($A_3^1$)}
\put(105,58){$\mu^-$ ($e^-$)}
\put(98,25){$\bar{\nu}_\mu$ ($\bar{\nu}_e$)}
\put(100,-5){$\nu_\tau$}
\put(137,28){$\tau^-$}
\put(170,5){$W^-$}
\put(250,-5){$\mu^-$ ($e^-$)}
\put(250,25){$\bar{\nu}_\mu$ ($\bar{\nu}_e$)}
\put(255,59){$\nu_\tau$}
  \includegraphics[height=.09\textheight]{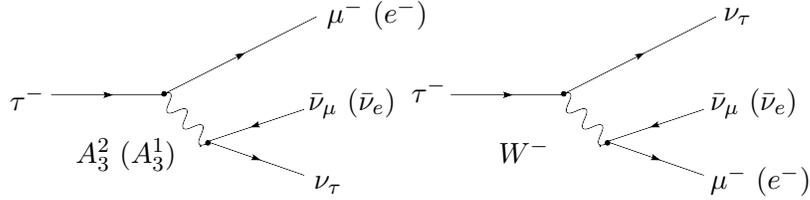}
  \hspace{15mm}
   \includegraphics[height=.09\textheight]{tau_decays.eps}
\end{picture}  
  \caption{Deviation from $e$-$\mu$ universality in tau decays}
  \label{tau_decay}
\end{figure}

We obtain a deviation from the $e$-$\mu$ universality
as follows:
\begin{equation}
R_\tau \equiv \frac{1+\epsilon_\mu}{1+\epsilon_e }
=\left[ 
\frac{B(\tau^- \rightarrow \mu^- \bar{\nu}_\mu \nu_\tau)}{
B(\tau^- \rightarrow e^- \bar{\nu}_e \nu_\tau)}
\frac{f(m_e/m_\tau)}{f(m_\mu/m_\tau)} \right]^{1/2} ,
\label{R_tau}
\end{equation}
where ${f(m_e/m_\tau)}/{f(m_\mu/m_\tau)}=1.028215$ and
\begin{equation}
\epsilon_\mu \simeq \frac{1}{4} z_2^2 r^2 = 1.4  \times 10^{-2} r^2,
\ \ \ 
\epsilon_e \simeq \frac{1}{4} z_1^2 r^2 = 6.8 \times 10^{-5} r^2 ,
\label{epsilon}
\end{equation}
with $r=v_W/v_\Psi$ ($v_W=246$ GeV).
Present experimental values \cite{PDG10} 
\begin{equation}
B(\tau^- \rightarrow \mu^- \bar{\nu}_\mu \nu_\tau )=
(17.39 \pm 0.04) \%, \ \ \  
B(\tau^- \rightarrow e^- \bar{\nu}_e \nu_\tau )=
(17.82 \pm 0.04) \% 
\label{br_exp}
\end{equation}
give 
\begin{equation}
R_\tau^{exp} = 1.0017 \pm 0.0016 , \ \ \ {\rm i.e.}
\  \epsilon_\mu \simeq 0.0017 \pm 0.0016 .
\label{Rtau_exp}
\end{equation}
This result seems to be in favor of the inverted 
gauge boson mass hierarchy although it is just at 1 $\sigma$ level.
(If the gauge boson masses take a normal hierarchy, 
$R_\tau$ will show $R_\tau < 1$.)
At present, we should not take the value \eqref{Rtau_exp} rigidly.
If we dare to adopt the center value in \eqref{Rtau_exp}, we
obtain $r\sim 0.35$ ($v_\Psi \sim 0.7$ TeV).
This value of $v_\Psi$ seems to be somewhat small.
We speculate $r\sim 10^{-1}$, i.e. $v_\Psi \sim$ a few TeV.   
The value will be confirmed by a tau factory in the near future.

\subsection{Family number conserved semileptonic decays of ps-mesons}

\begin{figure}
  \includegraphics[height=.3\textheight]{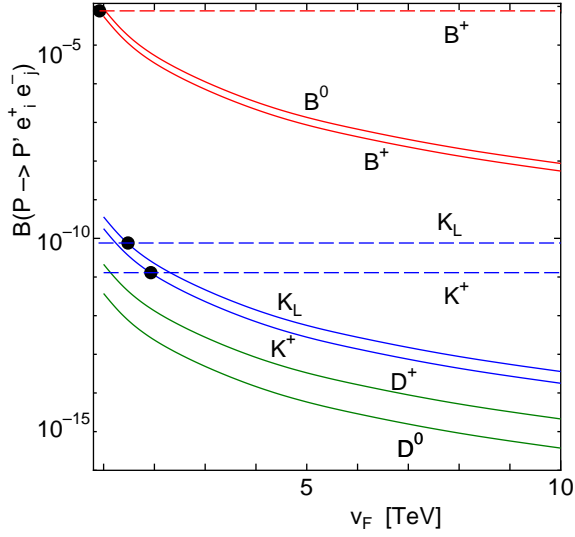}
  \caption{Predicted branching ratios $B(P \rightarrow P' e^+_i e^-_j)$
versus the VEV value $v_F \equiv v_{\Psi}$.
The marks $\bullet$ and the dashed lines denote present lower limits 
of the observed branching ratios. }
\label{psm-decay}
\end{figure}

Branching ratios of family number conserved semileptonic decays
are not sensitive to explicit values of the quark mixings $U_u$ 
and $U_d$.
We predict those in the limit of $U_u ={\bf 1}$ and $U_d={\bf 1}$:
\begin{equation}
\begin{array}{lll}
B(K^+\rightarrow \pi^+ \mu^+ e^-) & \sim 5 \times 10^{-12} & 
( < 1.3 \times 10^{-11}) , \\
B(K_L \rightarrow \pi^0 \mu^\pm e^\mp) & \sim 1 \times 10^{-11} & 
( <7.6 \times 10^{-11} ) , \\
B(D^+\rightarrow \pi^+ \mu^- e^+) & \sim 6 \times 10^{-13} & 
( <3.4 \times 10^{-5} ) , \\ 
B(D^0 \rightarrow \pi^0 \mu^- e^+) & \sim 1 \times 10^{-13} &
( <8.6 \times 10^{-5}) , \\ 
B(B^+\rightarrow K^+ \mu^- \tau^+) & \sim 2 \times 10^{-6} & 
( <7.7 \times 10^{-5} ) , \\ 
B(B^0\rightarrow K^0 \mu^- \tau^+) & \sim 2 \times 10^{-6} & 
( {\rm no\ data} ) . 
\end{array}
\label{ps-meson_decays}
\end{equation}
Here, the values in parentheses are present experimental upper 
limits \cite{PDG10}.
For reference, we illustrate the predicted branching ratios 
in Fig.~\ref{psm-decay}.
As seen in Fig.~\ref{psm-decay}, if $v_\Psi$ is a few TeV, 
observations 
of the lepton-flavor violating $K$- and $B$-decays with 
$\Delta N_F =0$ will be within our reach. 

\subsection{Direct searches at LHC and ILC}

\begin{figure}[h]
\begin{picture}(500,70)(-140,0)
\put(-15,52){$p$}
\put(-15,0){$p$}
\put(42,25){$A_3^3$}
\put(98,40){$b$}
\put(100,30){$\tau^-$}
\put(100,18){$\tau^+$}
\put(98,5){$\bar{b}$}
\put(193,30){$W^-$}
\put(240,40){$b$}
\put(245,30){$\tau^-$}
\put(245,18){$\tau^+$}
\put(245,5){$\bar{b}$}
  \includegraphics[height=.09\textheight]{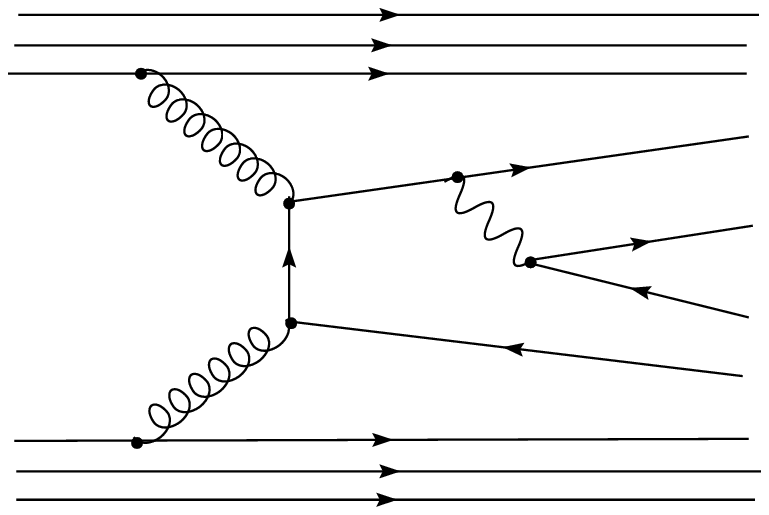}
  \hspace{15mm}
   \includegraphics[height=.09\textheight]{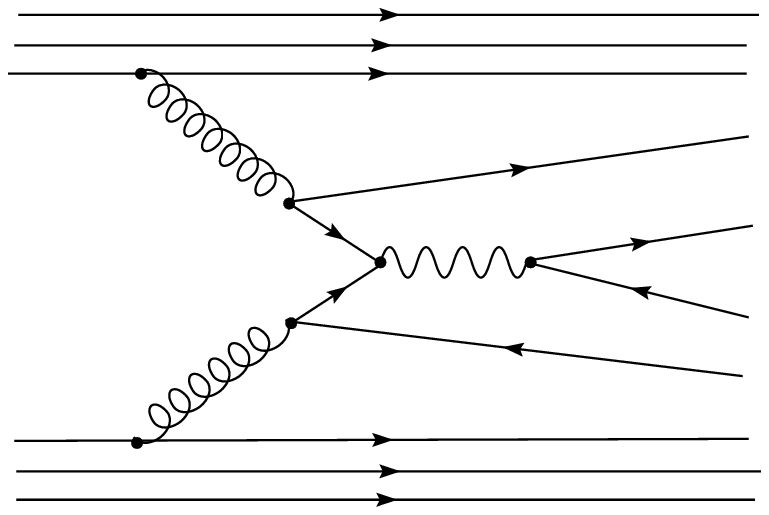}
\end{picture}  
  \caption{$A_3^3$ production at the LHC}
  \label{LHC}
\end{figure}

\begin{figure}[h]
\begin{picture}(500,70)(-120,0)
\put(-15,55){$e^-$}
\put(-15,-5){$e^+$}
\put(33,37){$Z^*$,$\gamma^*$}
\put(89,37){$A_3^3$}
\put(60,48){$\bar{b}$ ($\bar{t}$)}
\put(60,6){$b$ ($t$)}
\put(143,48){$\tau^-$}
\put(143,5){$\tau^+$}
\put(177,55){$e^-$}
\put(177,-5){$e^+$}
\put(225,35){$Z^*$,$\gamma^*$}
\put(265,39){$A_3^3$}
\put(280,55){$\bar{b}$ ($\bar{t}$)}
\put(280,-5){$b$ ($t$)}
\put(300,35){$\tau^-$}
\put(300,15){$\tau^+$}
  \includegraphics[height=.09\textheight]{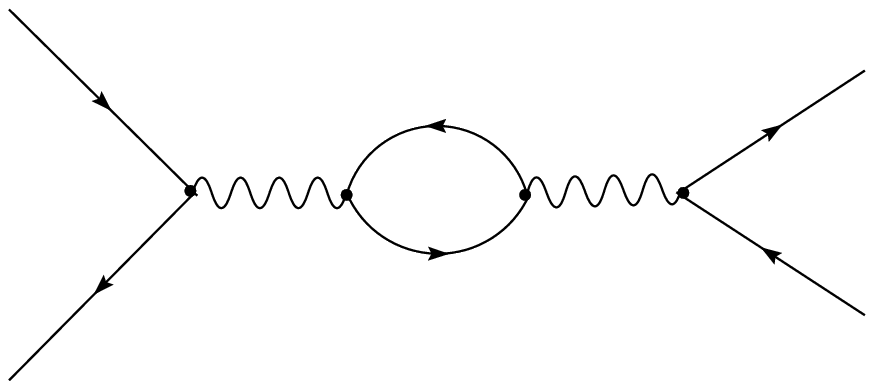}
  \hspace{15mm}
   \includegraphics[height=.09\textheight]{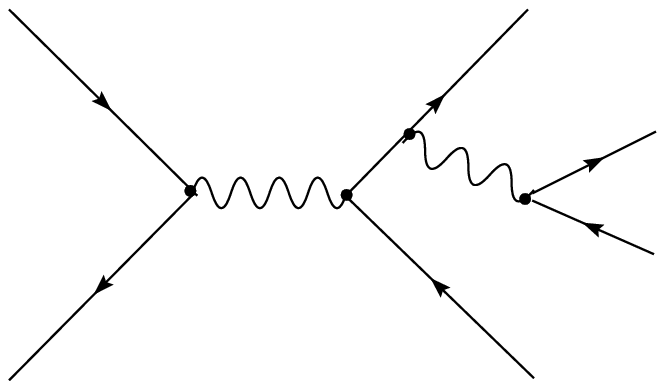}
\end{picture}  
  \caption{$A_3^3$ production at the ILC}
  \label{ILC}
\end{figure}

Since we speculate that $v_\Psi$ is of the order of a few TeV,
we may expect that the lightest family gauge boson $A_3^3$
is also of the order of a few TeV.
The experimental search is practically the same as that 
\cite{Z-prime} for $Z'$ boson,
but we will see only peak in $\tau^+\tau^-$ channel 
(no peaks in $e^+e^-$ and $\mu^+\mu^-$ channels). 
Possible productions of $A_3^3$ at the LHC and ILC are 
illustrated in Figs.~\ref{LHC} and \ref{ILC}, respectively.

\section{Summary}

The Sumino mechanism does not work for a SUSY model.
In order to work a Sumino-like mechanism for a SUSY model, we must
 consider a family gauge boson model with an inverted mass hierarchy.
Whether the gauge boson masses are inverted or normal is
 confirmed by observing the deviation form the $e$-$\mu$ universality in
 the pure leptonic tau decays, i.e. $R_\tau >1$ or $R_\tau <1$. 
The present observed values show $R_\tau^{exp}=1.0017\pm 0.0016$
 which is in favor of the inverted mass hierarchy.   
Since we speculate that the lightest gauge boson mass is a few TeV,
we expect the deviation $\Delta R_\tau \equiv R_\tau -1 \sim 10^{-4}$. 
A tau factory in the near future will confirm this deviation.
We also expect a direct observation of $\tau^+\tau^-$ in the LHC 
and the ILC. 

 \vspace{3mm}

{\Large\bf Acknowledgments} 

The authors would like to thank Y.~Sumino for valuable 
and helpful conversations, and also M.~Tanaka and K.~Tsumura for 
helpful discussions for lepton flavor violating and collider 
phenomenology.
One of the authors (YK) is supported by JSPS (No.\ 21540266).

\end{document}